\documentclass[prl,twocolumn,showpacs,amsmath,amssymb,superscriptaddress]{revtex4}
\usepackage{graphicx}
\begin{document}

\title{Analytically Solvable Model of Nonlinear Oscillations in a Cold but
Viscous and Resistive Plasma}

\author{E. Infeld}
\email[]{einfeld@fuw.edu.pl}
\affiliation{Department of Theoretical Physics, Soltan Institute for Nuclear
Studies, Ho\.za 69, 00--681 Warsaw, Poland}

\author{G. Rowlands}
\email[]{g.rowlands@warwick.ac.uk}
\affiliation{Centre for Fusion, Space and Astrophysics, Department of Physics,
University of Warwick, Coventry CV4 7AL, UK}

\author{A. A. Skorupski}
\email[]{askor@fuw.edu.pl}
\affiliation{Department of Theoretical Physics, Soltan Institute for Nuclear
Studies, Ho\.za 69, 00--681 Warsaw, Poland}

\date{\today}

\begin{abstract}
A method for solving model nonlinear equations describing plasma oscillations
in the presence of viscosity and resistivity is given. By first going to the
Lagrangian variables and then transforming the space variable
conveniently, the solution in parametric form is obtained. It involves
simple elementary functions. Our solution includes all known exact solutions
for an ideal cold plasma and a large class of new ones for a more realistic
plasma. A new nonlinear effect is found of splitting of the largest
density maximum, with a saddle point between the peaks so obtained.
The method may sometimes be useful where Inverse Scattering fails.
\end{abstract}

\maketitle

This paper concerns a fluid description of a plasma and should be easily
understandable to any Reader familiar with fluid dynamics.

Exact solutions involving realistic situations with nonlinear effects
in plasmas are few and far between. Here we present a 1D analytically solvable
model of nonlinear oscillations in a two-component viscous and resistive plasma.
We will obtain generalized plasma waves.

As is well known to plasma physicists, when viscosity and pressure are
neglected, a simple Lagrangian coordinate transformation leads to a known,
oscillating, nonlinear solution \cite{david}. Here, a second coordinate
transformation defined by the initial density profile will be required.

We assume that our plasma contains one kind of $Z=1$ ion (protons, deuterons
or tritons). They represent a uniform and motionless background for electron
fluid oscillations. The latter are mainly driven by the electric field $E(x)$.
Electron pressure forces will be neglected. Together with the assumed
infinitely heavy ions this eliminates ion-acoustic modes.
Denoting by $n_0$ ($= \text{const}$) the number density of ions, the basic
equations describing the electron fluid are: the continuity equation, momentum
transfer equation and Poisson's equation,
\begin{eqnarray} 
\frac{\partial n_e}{\partial t} + \frac{\partial}{\partial x} \bigl(
n_e v_e \bigr) &=& 0,\label{cont}\\
m n_e \Bigl( \frac{\partial v_e}{\partial t} + v_e \frac{\partial v_e}{\partial
x} \Bigr) &=& -\frac{\partial}{\partial x}(n_e T_e) - e n_e E\nonumber\\
&&+ \frac{\partial}{\partial x} \biggl( \frac{4}{3} \nu_e
\frac{\partial v_e}{\partial x} \biggr) - \eta \, e^2 n_0 n_e v_e,\nonumber\\
&&\label{mom}\\
\frac{\partial E}{\partial x} &=& 4 \pi e ( n_0 - n_e ),\label{pois}
\end{eqnarray}
where the electron temperature $T_e$ is in energy units, $\nu_e$ is the electron
viscosity coefficient and $\eta$ is the plasma resistivity.
We assume that $T_e$ is sufficiently small, see the following condition
(\ref{Tmax}), so that the first term on the right hand side in Eq.~(\ref{mom})
is negligible as compared to the second term. Furthermore, $\nu_e(x,t)$ will
be modelled as
\begin{equation}
\label{viscm} 
\frac{4}{3}\nu_e(x,t) = \nu\frac{n_0}{n_e(x,t)}, \quad \nu = \text{const}.
\end{equation}
This will allow us to solve Eqs.~(\ref{cont})--(\ref{pois}) analytically.

Analytical formulas for the electron viscosity coefficient $\nu_e(x,t)$ and
resistivity $\eta(x,t)$ involve several approximations, see e.g. \cite{brag}.
As a rule one assumes that the plasma is quasineutral ($n_e \approx n_i$) and
the distribution functions are not far from local Maxwellians. Other
approximations come from the Chappman--Ensgog method, where only first order
corrections to local Maxwellians are included and terms containing derivatives
are neglected. Therefore, uncertainty factors of two or three cannot be excluded.
In a $Z=1$ plasma, the relevant formulas given in the first reference
of \cite{brag} take the form:
\begin{subequations}
\label{trco}
\begin{equation}
\nu_e = 0.73 \frac{3 \sqrt{m} T_e^{5/2}}{4 \sqrt{2\pi} \lambda_Q e^4} =
4.0245\!\times\!10^{-8} \frac{T_e[eV]^{5/2}}{\lambda_Q/10},\label{nue}
\end{equation}
\begin{equation}
\eta = 0.51 \frac{4\sqrt{2\pi m} \lambda_Q e^2}{3 T_e^{3/2}} = 5.8524%
\!\times\!10^{-14} \frac{\lambda_Q/10}{T_e[eV]^{3/2}},\label{eta}
\end{equation}
\end{subequations}
where $\lambda_Q$ is a slowly varying Coulomb logarithm (typically $\lambda_Q
\approx$ 10--20). It can be seen that if $n_e/n_0$ is not much different from
unity and $T_e = \text{const}$, then both quasineutrality is approximately
valid and our modelling (\ref{viscm}) is approximately consistent with
Eq.~(\ref{nue}).

Using Eq.~(\ref{viscm}) and neglecting the electron pressure term, we can write
Eq.~(\ref{mom}) in the form
\begin{equation}
\frac{\partial v_e}{\partial t} + v_e \frac{\partial v_e}{\partial x} =
-\frac{e}{m}\,E + \frac{\nu}{m n_e} \frac{\partial}{\partial x}
\Bigl(\frac{n_0}{n_e}\frac{\partial v_e}{\partial x}\Bigr) - \eta \,
\frac{e^2 n_0}{m} \, v_e.
\label{veq} 
\end{equation} 

The first step to an analytical solution of Eqs.~(\ref{cont})--(\ref{pois}) is
to introduce Lagrangian coordinates: $x_0(x,t)$, the initial position (at $t=0$)
of an electron fluid element which at time $t$ was at $x$, and time in the
electron fluid rest frame, $\tau (= t)$. The basic transformation equations
between Eulerian and Lagrangian coordinates are (see \cite{infbook} for more
detail)
\begin{equation}\label{trans}
\frac{\partial}{\partial \tau} = \frac{\partial}{\partial t} + v_e
\frac{\partial}{\partial x}, \quad x = x_0 +
\int_0^{\tau}v_e(x_0,\tau')\,d\tau'.
\end{equation}
The continuity equation (\ref{cont}) in the electron fluid rest frame is simply
\begin{equation}\label{cont1}
n_e \frac{\partial x}{\partial x_0} = n_{e0}(x_0) \equiv n_e(x, t=0).
\end{equation}
Therefore, if we introduce an auxiliary variable $s$ related to $x_0$
by the 1D transformation:
\begin{equation}
\frac{ds}{dx_0} = \frac{n_{e0}(x_0)}{n_0}, \quad s(x_0 = 0) = 0, \label{svar} 
\end{equation}
we obtain
\begin{equation}
n_e(x,t) = \frac{n_{e0}(x)}{\dfrac{\partial x}{\partial x_0}} =
\frac{n_0}{\dfrac{\partial x}{\partial s}}.\label{ne} 
\end{equation}
Eq.~(\ref{ne}) is equivalent to
\begin{equation}
\frac{n_0}{n_e} \frac{\partial}{\partial x} =
\frac{\partial}{\partial s}, \quad \text{or} \quad ds = \frac{n_e}{n_0} \, dx.
\label{ds}
\end{equation}
Using Eqs.~(\ref{trans}) and (\ref{ds}), Eq.~(\ref{veq}) takes the linear form
\begin{equation}
\frac{\partial v_e}{\partial \tau} + \frac{e}{m}\,E - \frac{\nu}{m n_0}
\frac{\partial^2 v_e}{\partial s^2} + \eta \, \frac{e^2 n_0}{m} \, v_e= 0.
\label{veql} 
\end{equation} 
An important point is that $E$ can also be linearly expressed in terms
of $v_e$. Indeed, using
\begin{equation}
\label{dEdt}
\frac{\partial E}{\partial t} = 4\pi e n_e v_e ,
\end{equation}
which follows from Eqs.~(\ref{pois}) and (\ref{cont}), and adding this to
$v_e$ times Eq.~(\ref{pois}) we end up with
\begin{equation}
\label{dEdta} 
\frac{\partial E}{\partial\tau} = 4\pi e n_0 v_e ,
\end{equation}
where the right hand side is linear in $v_e$ as promised. Eqs.~(\ref{veql})
and (\ref{dEdta}) lead to
\begin{subequations}
\label{veql1}
\begin{equation} 
\frac{\partial^2 v_e}{\partial \tau^2} - \frac{\nu}{m n_0} 
\frac{\partial^2}{\partial s^2}\frac{\partial v_e}{\partial \tau}
+ \eta \, \frac{\omega_p^2}{4\pi} \frac{\partial v_e}{\partial\tau}
+ \omega_{p}^2 v_e = 0,
\end{equation} 
\begin{equation}
\omega_{p}^2 = \frac{4 \pi n_0 e^2}{m},
\end{equation}
\end{subequations}
which is a \textit{linear} partial differential equation for $v_e(s,\tau)$ with
\textit{constant} coefficients. Solutions of such PDEs are any superpositions
of the normal modes $\exp[i(ks - \omega\tau)]$, for which Eq.~(\ref{veql1})
leads to the dispersion relation
\begin{subequations}
\label{disp} 
\begin{eqnarray}
\omega &=& - i \alpha_k \pm \omega_k,\\
\alpha_k &=& \frac{\nu k^2}{2 m n_0} + \frac{\eta \, e^2 n_0}{2 m}, \quad
\omega_k = \sqrt{\omega_p^2 - \alpha_k^2}.
\end{eqnarray}
\end{subequations}
Assuming that $k$ is real and superposing the normal modes corresponding to
the plus and minus signs in $\omega$ given by Eq.~(\ref{disp}) we obtain four
real solutions:
\begin{equation} 
v_e = \exp(-\alpha_k\tau) f(\omega_k \tau) \, g(ks),\label{vsol}
\end{equation} 
where $f,g = \sin \: \text{or} \: \cos$. Our choice will be $f = g =\sin$, for
which $v_e(s,\tau = 0) \equiv 0$ and all three variables $x$, $\xi$ and $s$
will have a common origin ($x = \xi = s =0$).

For each $\tau$, $v_e$ given by Eq.~(\ref{vsol}) is a periodic function of $s$
with wavelength $\lambda=2\pi/k$. By adding higher harmonics, obtained from
Eq.~(\ref{vsol}) on replacing $k \to nk$, $n = 1,2,\ldots$, and multiplying by
an amplitude, any solution periodic in $s$ with wavelength $\lambda$ can be
obtained. Our choice will be
\begin{equation} 
v_e = \sum_{n=1}^{\infty} A_n \exp( -\alpha_{nk} \tau ) 
\sin(\omega_{nk} \tau) \sin(nks).\label{vsum}
\end{equation} 

Our equations and final results take a simple and universal form if we
introduce dimensionless quantities which will be denoted by a bar:
\begin{subequations}
\label{dimlpar}
\begin{eqnarray}
\bar{x} &=& kx, \quad \bar{s} = ks, \quad \bar{t} = \omega_p t, \quad
\bar{\tau} = \omega_p \tau,\\
\bar{\omega}_n &=& \frac{\omega_{nk}}{\omega_p} = \sqrt{1 - \bar{\alpha}_n^2},
\quad  \bar{\alpha}_n = n^2 \bar{\nu} + \bar{\eta},\\
\bar{\eta} &=& \frac{\eta \, n_0 e^2}{2 m\omega_p} = 1.3136\!\times\! 10^{-10}
\frac{\lambda_Q}{10} \frac{n_0^{1/2}}{T_e[eV]^{3/2}},\label{etab}\\
\bar{\nu} &=& \frac{2\nu_e k^2}{3 m n_0 \omega_p} = 2.7076\!\times\! 10^6
\frac{T_e[eV]}{\bar{\eta} n_0 \lambda^2},\label{nub}\\
\bar{v}_e &=& \frac{v_e}{v_{\text{ph}}}, \quad \bar{A}_n =
\frac{A_n \bar{\omega}_n}{v_{\text{ph}}}, \quad  v_{\text{ph}} =
\frac{\omega_p}{k},\\
\bar{n}_e &=& n_e/n_0, \quad \bar{E} = E e k/(m\omega_p^2).
\end{eqnarray}
\end{subequations}
Thus, Eqs.~(\ref{pois}), (\ref{ne}) and (\ref{dEdta}) are:
\begin{equation}
\frac{\partial\bar{E}}{\partial\bar{x}} = 1 - \bar{n}_e, \quad
\frac{\partial\bar{s}}{\partial\bar{x}} = \bar{n}_e, \quad
\frac{\partial\bar{E}}{\partial\bar{\tau}} = \bar{v}_e.\label{dimleqs}
\end{equation}
Integrating the first two over $d \bar{x}$ and the last one over
$d \bar{\tau}$ and using (\ref{vsum}), we end up with equations
which define all relevant quantities: $\bar{x}$, $\bar{E}$,
$\bar{n}_e$ and $\bar{v}_e$ in terms of $\bar{s}$ and $\bar{\tau}$ ($=\bar{t}$).
Dropping bars for simplicity, the final results for the dimensionless quantities
become:
\begin{subequations}
\label{dimles}
\begin{eqnarray}
x &=& s + E ,\label{barx}\\
E(s,t) &=& - \sum_{n=1}^{\infty} A_n
G_n(t) \sin(n s),\label{barE}\\
n_e(s,t)^{-1} &=& 1 - \sum_{n=1}^{\infty} n A_n
G_n(t)\cos(n s),\label{barne}
\end{eqnarray}
\begin{equation}
v_e(s,t) = \sum_{n=1}^{\infty} A_n \exp(- \alpha_n t)
\frac{\sin(\omega_n t)}{\omega_n} \sin(n s),
\end{equation}
\begin{equation}
G_n(t) = \exp(- \alpha_n t)\Bigl[
\alpha_n \frac{\sin(\omega_n t)}{\omega_n}
+ \cos(\omega_n t) \Bigr],
\end{equation}
\end{subequations} 
where one should replace $\sin(\omega_n t)/\omega_n$ by
$t$ if $\omega_n = 0$. One can notice that if $n^2\nu + \eta
> 1$, $\omega_n (= i \lvert\omega_n\rvert)$ becomes pure imaginary
but $G_n(t)$ remains real ($\sin(\omega_n t)/
\omega_n = \sinh(\lvert\omega_n\rvert t)/\lvert
\omega_n\rvert$, and $\cos(\omega_n t) =
\cosh(\lvert\omega_n\rvert t)$.

Eq.~(\ref{barne}) is only meaningful if the sum is
smaller than unity, which imposes a limitation on the $A_n$.

\begin{figure}[t]
\centerline{\includegraphics[scale=.34]{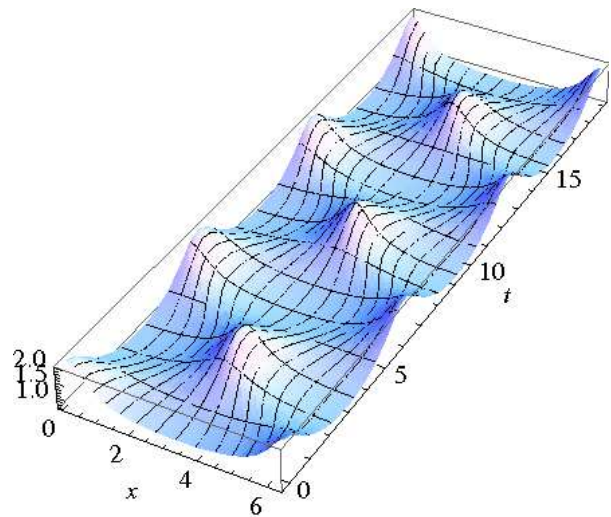}}
\caption{(color online). Plot of $n_e(x,t)$ when only
 $A_1 = 0.5$ is nonzero, and $\nu = \eta =10^{-5}$. This corresponds to
 $n_0 = 1.45\times10^{18}$ m$^{-3}$ and $\lambda = 4.32$ m, if we assume
 $T_e = 10$ eV and $\lambda_Q/10 = 2$. Damping is practically invisible for the 
 times presented.}\label{fig1}
\end{figure}

\begin{figure}[h!]
\centerline{\includegraphics[scale=.36]{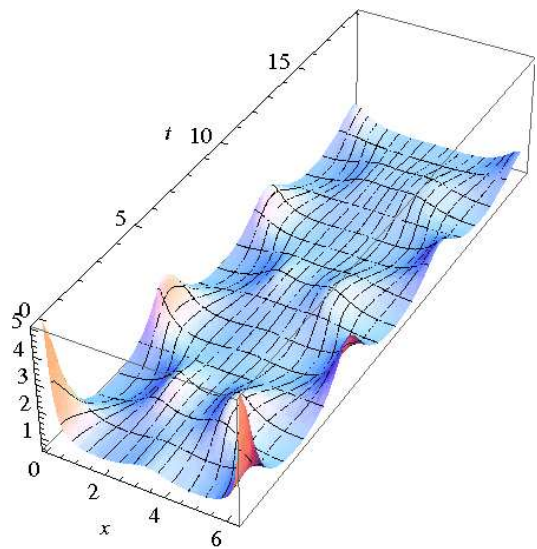}}
\caption{(color online). Plot of $n_e(x,t)$ when only
$A_1 = 0.4$ and $A_2 = 0.2$ are nonzero, $\nu = 0.02$ and $\eta =
10^{-6}$. This corresponds to $n_0 = 1.45\times10^{13}$ m$^{-3}$ and $\lambda =
30.6$ m, if we assume $T_e = 1$ eV and $\lambda_Q/10 = 2$. The viscous damping
is evident after a single period.}\label{fig2}
\end{figure}

\begin{figure}[t!]
\centerline{\includegraphics[scale=.39]{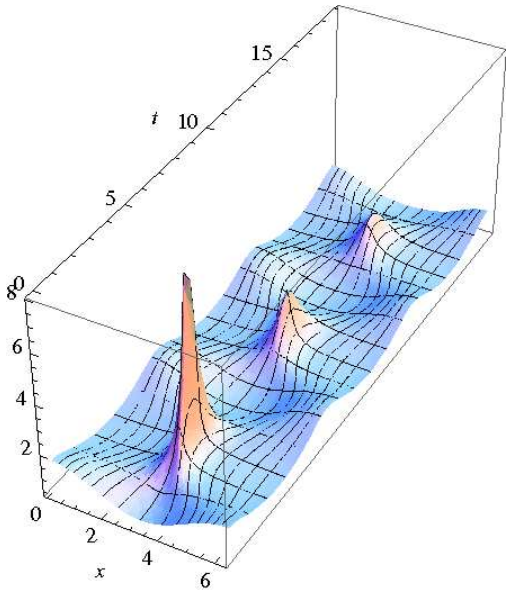}}
\caption{(color online). Plot of $n_e(x,t)$ when only
$B_1(0) = 0.55$ is nonzero, $\nu = 0.015$, $\eta = 10^{-5}$ and $N=20$. This
corresponds to $n_0 = 1.45\times10^{21}$ m$^{-3}$ and $\lambda = 11.2\times
10^{-3}$ m, if we assume $T_e = 100$ eV and $\lambda_Q/10 = 2$.
Note that in the presence of even weak viscosity $B_1(0)$ can exceed $1/2$,
see Eqs.~(\ref{dimles}).
}
\label{fig3}
\end{figure}

\begin{figure}[h!]
\centerline{\includegraphics[scale=.35]{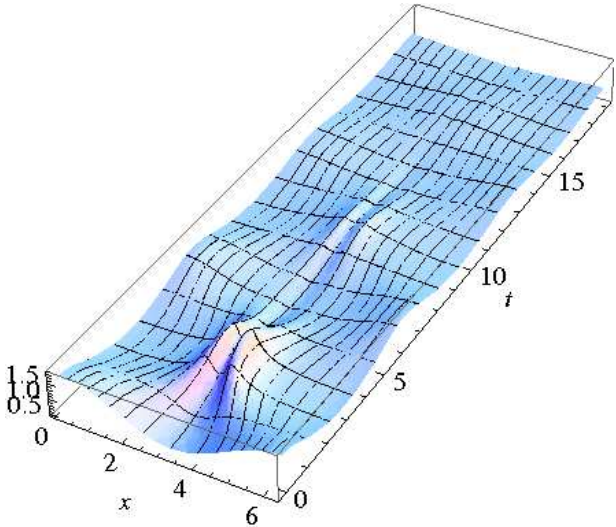}}
\caption{(color online). Plot of $n_e(x,t)$ when only
$B_1(0) = 0.55$ is nonzero, $\nu = 0.1$ $\eta = 10^{-4}$ and $N=20$. This
corresponds to $n_0 = 1.45\times10^{20}$ m$^{-3}$ and $\lambda = 1.36\times
10^{-3}$ m, if we assume $T_e = 10$ eV and $\lambda_Q/10 = 2$.
The observed bifurcation of the maximum under strong viscosity is a new
nonlinear effect.
}\label{fig4}
\end{figure}

One can eliminate the parameter $s$ from Eqs.~(\ref{dimles}) thereby making $x$
and $t$ the independent variables. This parameter has to be determined in terms
of $x$ and $t$ from Eq.~(\ref{barx}) and used in  the remaining
Eqs.~(\ref{dimles}). While numerically this is a simple task, analytical
formulas are complicated (though for $s = 0$ or $\pi$, we obtain simply
$s = x$). At the same time, the parametric form (\ref{dimles}), which involves
simple elementary functions, can also be used to plot $E(x,t)$, $n_e(x,t)$ and
$v_e(x,t)$ (e.g., by using ParametricPlot3D of Mathematica, see
Figs.~\ref{fig1}--\ref{fig4}).

We can expect our solution (\ref{dimles}) to be realistic if $n_e(x,t)$
is not much different from unity. Under this assumption, the electron pressure
term in Eq.~(\ref{mom}) is negligible as compared to the electric field if
$T_e \ll m v_{\text{ph}}^2 \equiv e^2 n_0 \lambda^2/\pi$, equivalent to 
\begin{equation}
T_e[eV] \ll 4.5835 \times 10^{-8} n_0 \lambda^2 \quad \text{or} \quad
\nu\,\eta \ll 0.124.\label{Tmax}
\end{equation}
This indicates that at least one of the coefficients $\nu$ or $\eta$ must be
much smaller than $0.3$. One can prescribe these coefficients and one of the
parameters $T_e[eV]$, or $n_0$  and determine the second one from
Eq.~(\ref{etab}) and then find $\lambda$ from Eq.~(\ref{nub}).

In Figs.~\ref{fig1} and \ref{fig2} we present typical examples for a small
number of modes included. Various spatially periodic structures can be
produced. Note the relevance of our considerations to a wide range of both
laboratory and space plasmas.

The electron density $n_e(x,t)$ is an even and periodic function of $x$ with
wavelength $2\pi$. It can be expanded in a Fourier series in $x$ with time
dependent Fourier coefficients:
\begin{equation}
\label{nefx}
n_e(x,t) = 1 + \sum_{m=1}^{\infty} B_m(t) \cos(m x).
\end{equation}
This along with the first equation in (\ref{dimleqs}) integrated over
$dx$ leads to the Fourier expansion of $E(x,t)$:
\begin{equation}
\label{Efx}
E(x,t) = - \sum_{m=1}^{\infty} \frac{B_m(t)}{m} \sin(m x).
\end{equation}
And finally, Eqs.~(\ref{dEdt}) and (\ref{Efx}) lead to
\begin{equation}
\label{neve}
n_e(x,t) v_e(x,t) =
\frac{\partial E}{\partial t} = - \sum_{m=1}^{\infty}
\frac{dB_m(t)}{dt} \frac{\sin(mx)}{m},
\end{equation}
from which $v_e(x,t)$ can be determined.

The coefficients $A_n$ and $B_m(t)$ are related by
\begin{equation}
\label{AnBm}
\begin{split}
A_n& = -\frac{2}{n\pi} \int_0^{\pi} \cos\Bigl\{ n \Bigl[x +
\sum_{m=1}^{\infty} \frac{B_m(0)}{m}\sin(mx) \Bigr] \Bigr\} \, dx,\\
B_m(t)& = \frac{2}{\pi} \int_0^{\pi} \cos\Bigl\{ m \Bigl[s -
\sum_{n=1}^{\infty} A_n G_n(t)\sin(ns) \Bigr] \Bigr\} \, ds.
\end{split}
\end{equation}
These integrals are not expressible in terms of elementary functions, but if
the sum over $m$ or $n$ is truncated at some $M$ or $N$, one can easily
calculate numerically as many integrals as needed. If either $M=1$ or $N=1$,
$A_n$ or $B_m(t)$ is expressible in terms of Bessel functions. Thus using the
identity \cite{rizik} (p. 185)
\begin{equation}
\int_0^{\pi} \cos(ax - z\sin x)\,dx = \pi J_a(z),\label{ident}
\end{equation}
we obtain respectively
\begin{equation}
A_n = \frac{2}{n} (-1)^{n+1} J_n(n B_1(0)) \quad \text{if }
B_m(0) = 0 \text{ for } m > 1,\label{Anp}
\end{equation}
\begin{equation}
B_m(t) = 2 J_m(m A_1 G_1(t)) \quad \text{if }
A_n = 0 \text{ for } n > 1.\label{Bmp}\end{equation}
Eqs.~(\ref{AnBm}) result from standard formulas for Fourier coefficients,
for the function $n_e(s,t=0)^{-1}$ or $n_e(x,t)$ as a function of $x$, if the
integration variable in these formulas is changed from $s = x - E$ to $x$
($n_e^{-1}\,ds = dx$) or conversely from $x = s + E$ to $s$ ($n_e\,dx = ds$),
see Eqs.~(\ref{barx}) and (\ref{ds}).

Eqs.~(\ref{Bmp}) and (\ref{nefx})--(\ref{neve}) present a new solution
explicitly given in terms of physical variables $x$ and $t$. Its plot for
$\nu = \eta =10^{-5}$ is shown in Fig.~\ref{fig1}.

The particular solution (\ref{dimles}) with $A_n$ given by (\ref{Anp})
reduces to that of \cite{david} if $\nu = \eta = 0$ though in a different
notation. The known condition $B_1(0) < 1/2$ follows from the identity
$\sum_{n=1}^{\infty} J_n(n\alpha) = \frac{\alpha}{2(1 - \alpha)}$
\cite{rizik} (p. 366). For $B_1(0) = 1/2$, $n_e^{-1}$
becomes zero at $s = t = \pi$.

The behavior of this solution for $\nu,\eta > 0$ is shown in Figs.~\ref{fig3}
and \ref{fig4}. In a viscous and resistive plasma
$B_1(0)$ can exceed $1/2$. Furthermore if $\nu$ is sufficiently
large, see Fig.~\ref{fig4}, where $\nu = 0.1$, a new nonlinear effect can
be noticed, i.e., the largest density maximum splits in two, with a saddle
point between the peaks.

\end{document}